\documentclass{PoS}

\usepackage[utf8]{inputenc}
\usepackage{mathtools}
\usepackage{amsfonts}
\usepackage{amssymb}
\usepackage{fontenc}
\usepackage{times}
\usepackage{mathptmx}
\usepackage{graphicx}
\usepackage{slashed}
\usepackage{float}

\newcommand{\overbar}[1]{\mkern 1.5mu\overline{\mkern-1.5mu#1\mkern-1.5mu}\mkern 1.5mu}

\newcommand{\EQ}{Eq.~\ref}

\usepackage{subcaption}

\usepackage{textpos}

\usepackage[toc,page]{appendix}
\usepackage{pdfpages}

\title{Constraining gluon PDFs with quarkonium production}

\ShortTitle{Constraining gluon PDFs}

\author{\speaker{Melih~Arslan~Ozcelik}\\
Institut de Physique Nucl\'eaire Orsay, CNRS/IN2P3, Univ. Paris-Sud, Université Paris-Saclay, Orsay, France\\
E-mail: \email{ozcelik@ipno.in2p3.fr}}

\abstract{We explore how the positivity of the $P_T$-integrated $\eta_c$-hadroproduction cross-section computed at NLO in $\alpha_s$ can set up constraints on the $x$-dependence of gluon PDFs at low scales.
}

\FullConference{XXVII International Workshop on Deep-Inelastic Scattering and Related Subjects - DIS2019\\
		8-12 April, 2019\\
		Torino, Italy}

\begin{document}

\section{Introduction}

In these proceedings, we will discuss how the $P_T$-integrated $\eta_c$ hadro-production cross-section at NLO in $\alpha_s$ can set up constraints on the $x$-dependence of gluon PDFs at low scales. From the theoretical point of view, $\eta_c$ ($^1S_0$) is the simplest of all quarkonia to deal with and can reveal much about the bound-state formalism. Throughout the proceedings, in the framework of Non-Relativistic QCD (NRQCD), we will be dealing with the colour-singlet case, which is the leading contribution in the $v^2$-expansion of NRQCD. For reviews on quarkonium production, we guide the readers to Refs.~\cite{Lansberg:2019adr, Andronic:2015wma, Brambilla:2010cs, Lansberg:2006dh}.

PDFs (Parton Distribution Functions) depend on both its parton momentum fraction $x$ and the factorisation scale $\mu_F$. Different PDF sets use different parametrisations and data in their fits. They are typically parametrised at a scale around the mass of the charm quark. At higher factorisation scales as in Higgs production, it is the QCD evolutions governed by the DGLAP equations that will become relevant. At these scales the gluon PDFs do not depend much on the initial parametrisations anymore. They become almost identical. For low-scale processes however, hadronic cross-sections will be particularly sensitive to the initial parametrisation. Therefore we advocate the use of $\eta_c$ as a probe to study the $\sigma_{hh}$ dependence on the different PDF parametrisations.

It was the problem of negative cross-section that led us to conduct this study here. We will go into more detail in the next sections and show that the cross-section dependence on the renormalisation and factorisation scales depends much on the choice of the PDF. We will also elaborate on what has been done in this field in the past and what the prospects for the future are.

\section{$\eta_c$ production and the issue of negative cross-sections}

The $\eta_c$ is a gluon probe at low scales and the simplest of all quarkonia as far as the computation of hadro-production cross-sections is concerned. The transverse-momentum $P_T$- and the rapidity $y$-integrated cross-section is known at NLO in $\alpha_s$ since 1992 in collinear factorisation \cite{Kuhn:1992qw}.
The first hadro-production measurement data was released only recently in 2015 by the LHCb collaboration for $P_T\geq 6$~GeV at $\sqrt{s}=7$ and $8$ TeV \cite{Aaij:2014bga}. As can be seen in Fig.~2 of Ref.~\cite{Butenschoen:2014dra}, the NLO Colour-Singlet Model works well for $\eta_c$ \cite{Han:2014jya, Zhang:2014ybe}. The data set does not cover the low-$P_T$ region, it could however be measured down to $P_T=0$ using the LHC beams in the fixed-target mode, generically called AFTER@LHC \cite{Hadjidakis:2018ifr, Feng:2019zmn}. For $\eta_c$ and other low scale quarkonia bound-states we encounter the issue of \textit{negative} cross-sections in perturbative calculations.

\begin{figure}[h]
	\begin{subfigure}{0.5\textwidth}
		\captionsetup{oneside,skip=13pt}
		\centering
		\includegraphics[scale=0.30]{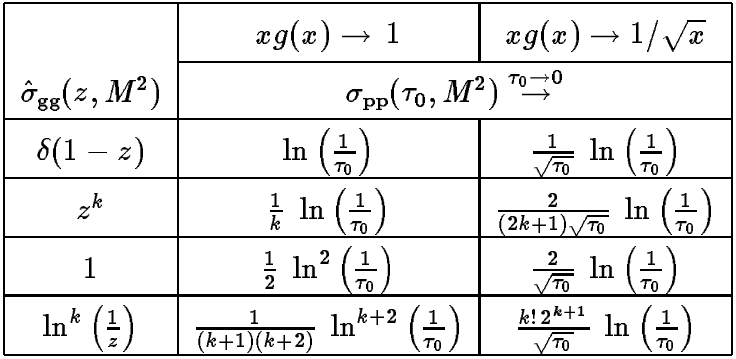}
		\caption{Asymptotic ($\tau_0=M^2/s\rightarrow 0$) behaviour of the proton-proton or proton-antiproton cross section for various terms of the gluon-gluon subprocess ($z=M^2/\hat{s}=\tau_0/\tau$) and two extreme choices of the gluon distribution function. Taken from Schuler's Review \cite{Schuler:1994hy}, 1994.}
		\label{fig:schulertable}
	\end{subfigure}
	\quad
	\begin{subfigure}{0.47\textwidth}
			\captionsetup{oneside,margin={0.7cm,0cm}}
			\captionsetup{skip=-1pt}
			\centering
			\includegraphics[scale=1.2]{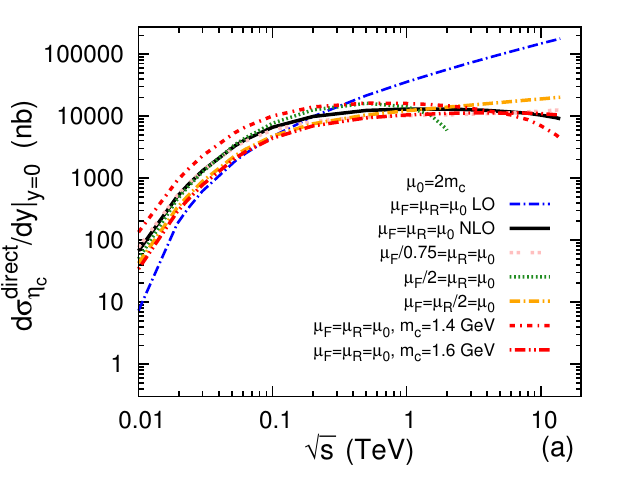}
			\caption{comparison of $\eta_c$ differential cross-section at NLO with different choices of $\mu_R$ and $\mu_F$ with CTEQ6M, taken from Ref.~\cite{Feng:2015cba}}
			\label{fig:etaccomp}
	\end{subfigure}
\captionsetup{skip=-3pt}
\captionsetup{font=small,labelfont={bf}}
\setlength{\belowcaptionskip}{-20pt}
\caption{}
\end{figure}

As an example, we show in Fig.~\ref{fig:etaccomp} a plot from Ref.~\cite{Feng:2015cba} for the rapidity-differential cross-section of $\eta_c$ at central rapidity $y=0$ and as a function of hadronic energy $\sqrt{s}$ for different scale choices. For some scale choices, the rapidity-differential cross-section becomes negative at large hadronic energies.

What are potential sources for \textit{negative} cross-sections? Is it due to failure of theoretical frameworks (NRQCD etc.) used to describe quarkonium production? Is it due to truncation intrinsic to fixed-order calculations? Do we need to go to higher orders (N$^2$LO, N$^3$LO, ...) to solve this issue?

To address these questions, it may be instructive to have a look into the case of open $c\overbar{c}$ production at NLO/N$^2$LO. In Ref.~\cite{Accardi:2016ndt}, the plots from Figs.~16 \& 17 show the $P_T$- and $y$-integrated cross-section as a function of $\sqrt{s}$ for different PDF sets at the default scale choice $\mu_R=\mu_F=2m_c$. What is remarkable is that for some PDF sets the cross-section can be negative at large $\sqrt{s}$. We are thus tempted to say that the issue of $\sigma\leq 0$ is not specific to the way quarkonium production is treated, but a more general one that involves low-scale physics. Concerning the truncation in fixed-order calculations, the plots demonstrate that when negative cross-sections occur already at NLO, the situation becomes even worse at N$^2$LO, hence it is not a problem of truncating an infinite series to a fixed-order. At this stage, we would like to stress that in Ref.~\cite{Accardi:2016ndt}, the authors attribute the issue of negative cross-sections to negative gluon PDFs at low scales and rather low-$x$ region, however the differential cross-section $d\sigma/dy$ is not yet available at N$^2$LO and a full scale analysis has not yet been performed. Therefore one cannot rule out the possibility of negative cross-sections with positive PDFs for open $c\overbar{c}$ production.

We may also want to consider whether the issue could be related to collinear factorisation itself. Do we need to include resummations of $\log{P_T}$? Is it due to an improper choice of renormalisation $\mu_R$ and factorisation $\mu_F$ scales? Or is it related to Parton Distribution Functions (PDFs)? We will address these questions in the next sections.

\section{Collinear factorisation \& PDF parametrisation}

In collinear factorisation, the cross-section for hadron-collision can be written as,
\begin{equation}
\sigma_{pp}=\sum_{ij} \int dx_1 dx_2\ f_{i/p}(x_1,\mu_F)f_{j/p}(x_2,\mu_F)\ \hat{\sigma}_{ij}(\mu_R,\mu_F,x_1,x_2,\hat{s}=s\, x_1 x_2),
\end{equation}
where $f_{i/p}(x_1,\mu_F)$ are the PDFs and $\hat{\sigma}_{ij}$ is the partonic cross-section for the relevant channel $ij$ where $i$ and $j$ are the partons. The hadronic cross-section has a dependence on both $\mu_R$ and $\mu_F$. In particular, the $\mu_R$-dependence implicitly comes from the strong coupling constant $\alpha_s$ and explicitly from ratios of dimensionful parameters at higher orders that origin from dimensional regularisation. The $\mu_F$-dependence occurs in both PDFs and explicitly inside the partonic cross-section via the Altarelli-Parisi terms, which are the counterterms to absorb the remaining collinear divergences into the PDFs.

We will now give a brief historical summary on what has been done in this field within collinear factorisation in the past. In 1992, K\"uhn \& Mirkes \cite{Kuhn:1992qw} computed the pseudo-scalar toponium cross-section at NLO. In 1994, G. Schuler published a Review \cite{Schuler:1994hy} in which he confirmed the result by K\"uhn \& Mirkes and he pointed out for the first time issues with negative cross-sections for quarkonia at large $\sqrt{s}$. He found an explanation why for some PDF choices there is a strong/weak scale dependence. Mangano and Petrelli in their 1996 Proceedings \cite{Mangano:1996kg} arrived to the same conclusions. The result by K\"uhn \& Mirkes was confirmed independently by Petrelli et \textit{al} \cite{Petrelli:1997ge} in 1997.

\subsection{Partonic high-energy limit}

In his Review, Schuler identified two potential sources of \textit{negative} cross-sections, the small-$x$ behaviour of the gluon and sea-quark distributions and the behaviour of the partonic cross-section far from threshold, where threshold is the limit $z=M^2/\hat{s}\rightarrow 1$. Both Schuler and Mangano found that the partonic cross-section at NLO in the partonic high-energy limit $z=M^2/\hat{s}\rightarrow 0$ has the general structure,
\begin{equation}
\begin{split}
\lim_{z\to 0}{\hat{\sigma}}_{gg}=2C_A\frac{\alpha_s}{\pi}\hat{\sigma}_{\textrm{Born}}\left(\log{\frac{M^2}{\mu_F^2}}-C_J\right), \quad
\lim_{z\to 0}{\hat{\sigma}}_{qg}=C_F\frac{\alpha_s}{\pi}\hat{\sigma}_{\textrm{Born}}\left(\log{\frac{M^2}{\mu_F^2}}-C_J\right),
\label{eq:Schulerlimit}
\end{split}
\end{equation}
where $C_J$ is a \textit{process-dependent} quantity and $M=2m_c$. The term $\log{M^2/\mu_F^2}$ is a \textit{universal} factor that originates from the Altarelli-Parisi counterterms. $C_J$ however is a quantity that uniquely comes from the real corrections and hence depends on the specific process. In the case of $\eta_c$ we have that $C_J=1$, for $\chi_c (^3P_{0,2})$ we have respectively $C_0=43/27$ and $C_2=53/36$. We would like to stress at this stage that $\lim_{z\to 0}{\hat{\sigma}}$ is particularly sensitive to the factorisation scale $\mu_F$. For the $\eta_c$, $C_J=1$ indicates that the cross-section at this limit is already negative for $\mu_F=M$ which is admittedly a reasonable scale choice. Variations from this scale will either make the limit more negative or in some cases positive. In fixed-order calculations, we however expect the result to be weakly dependent on $\mu_R$ and $\mu_F$ only. As a side note, we point out that the ratio of the $qg$- to the $gg$-channel approaches the value $C_F/(2C_A)=2/9$ for $z\rightarrow 0$. Note that this ratio is process-independent, hence it is the same for both bound and open $c\overbar{c}$ production and Higgs production with finite $m_t$ as well.

As it will be clear in a moment, \EQ{eq:Schulerlimit} allows us to understand the behaviour of the hadronic high-energy limit. As a simple toy model for the gluon PDFs $g(x)$, Schuler considered the functions $g(x)=1/x$ and $g(x)=1/x^{1.5}$. The table (see Fig.~\ref{fig:schulertable}) taken from his Review \cite{Schuler:1994hy}, shows how different partonic terms ($\delta\left(1-z\right)$, $z^k$, ...) of $\hat{\sigma}_{gg}$ translate to hadronic ones for both these PDFs.

From Fig.~\ref{fig:schulertable} we can conclude that with $g(x)=1/x$, the constant terms scale stronger with energy $\sqrt{s}$ than the terms $\delta(1-z)$ and $z^k$ at very large energies $\sqrt{s}\rightarrow \infty$. This brings us back to the partonic high-energy limit where it is precisely the constant terms of $\hat{\sigma}$ that survive in \EQ{eq:Schulerlimit} with $z\rightarrow 0$. Therefore with this 'flat' gluon PDF, the hadronic cross-section is sensitive to the factorisation scale $\mu_F$ and in particular to the sign of $\lim_{z\to 0}{\hat{\sigma}}$. This ultimately determines the behaviour of the cross-section at large $\sqrt{s}$.

For the second extreme parametrisation $1/x^{1.5}$ however, all partonic terms scale in the same way at large $\sqrt{s}$, therefore real corrections scale in the same way as threshold contributions (LO plus virtual corrections). The sensitivity on the factorisation scale $\mu_F$ is now diminished. Consequently, steeper gluon PDFs will damp down the real corrections which result into the dominance of threshold contributions at large $\sqrt{s}$. The NLO yield will therefore follow the shape of the LO cross-section.
We can conclude this section by stating that due to the low scale process, the cross-section crucially depends on the initial PDF parametrisation.

\vspace*{-0.25cm}
\section{$K$-factor for different PDFs}
\vspace*{-0.25cm}

Following our discussion on the sources of negative cross-sections, we will now illustrate in this section how the $\eta_c$ hadronic cross-section can depend strongly on the initial PDF parametrisation. In the first part of this section, we will display the $K$-factor, which is the ratio of the NLO cross-section over the LO one, of the $\eta_c$ cross-section differential at central rapidity $y=0$. In the second part, we will then briefly discuss the shape in rapidity $y$ of the cross-section.

For the $K$-factor we will consider five different PDF parametrisations \cite{Dulat:2015mca, Ball:2017otu, Martin:1995ws} (we will make use of the abbrevations in brackets), CT14nlo\_NF3 (CT14), NNPDF31sx\_nlo\_as\_0118 (NNPDFsx), NNPDF31sx\_nlonllx\_as\_0118 (NNPDFsxNLL), MRS(A'), MRS(G).
In order to differentiate between the PDF choices, we will make use of two different scale configurations, $\mu_R=\mu_F=2m_c=3$GeV and $\mu_R=m_c=1.5$GeV with $\mu_F=2m_c=3$GeV. By lowering the renormalisation scale $\mu_R$ in the second configuration, we are enhancing QCD corrections. The objective is to see the impact of the different gluon PDFs on the real corrections and, in particular, the constant terms that survive in the partonic high-energy limit (see \EQ{eq:Schulerlimit}).

\begin{textblock}{5}(8.9,1.15)
	\tiny MRS(G), $g(x)\sim 1/x^{1.30}$
\textblockcolour{white}
\end{textblock}
\begin{textblock}{5}(9.0,1.7)
	\tiny MRS(A'), $g(x)\sim 1/x^{1.14}$
\textblockcolour{white}
\end{textblock}
\begin{textblock}{5}(3.1,0.95)
	\tiny MRS(G), $g(x)\sim 1/x^{1.30}$
\textblockcolour{white}
\end{textblock}
\begin{textblock}{5}(3.2,1.45)
	\tiny MRS(A'), $g(x)\sim 1/x^{1.14}$
\textblockcolour{white}
\end{textblock}
\begin{figure}
	\centering
	\captionsetup{font=small,labelfont={bf}}
	\includegraphics[scale=0.307]{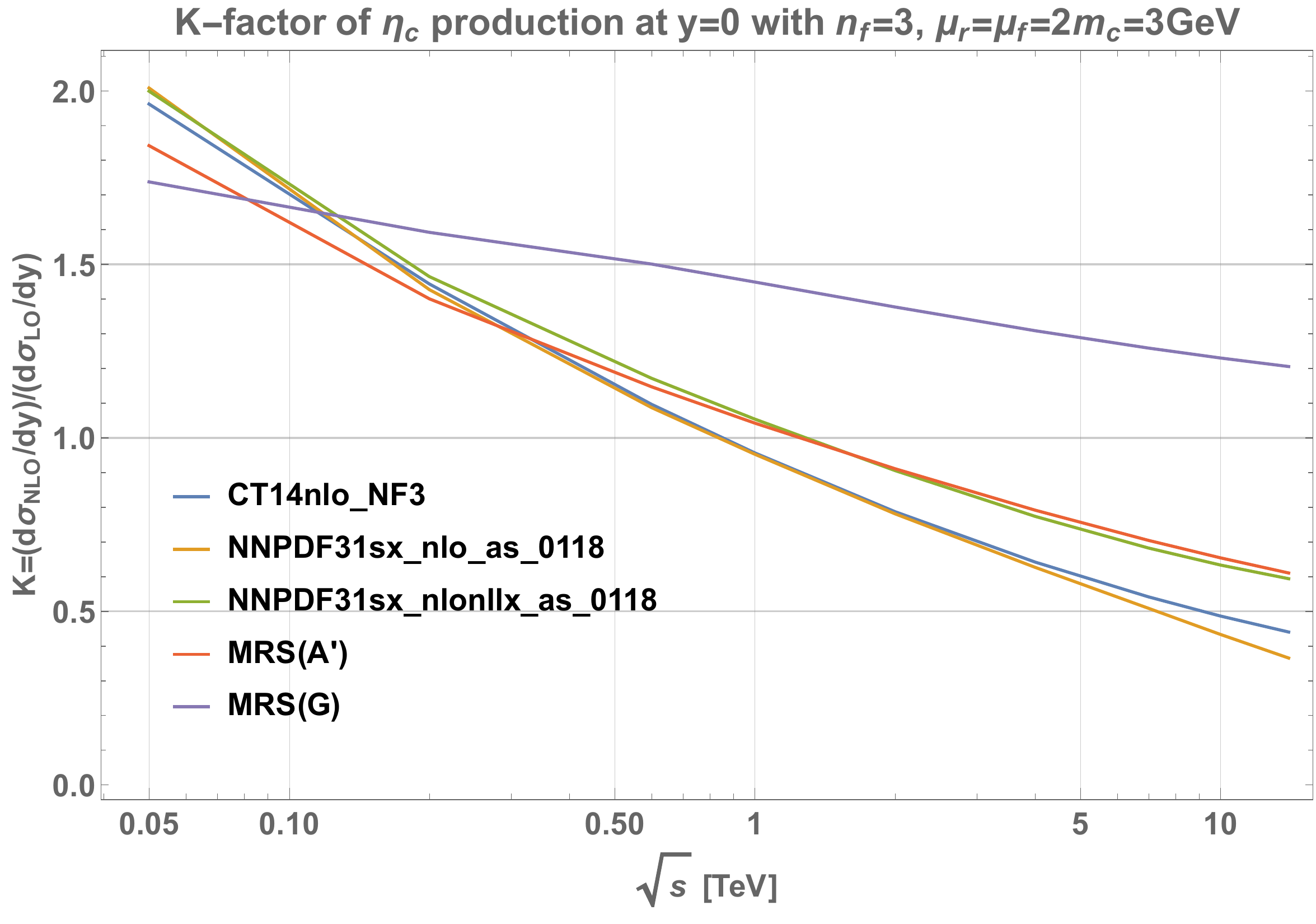}			\includegraphics[scale=0.310]{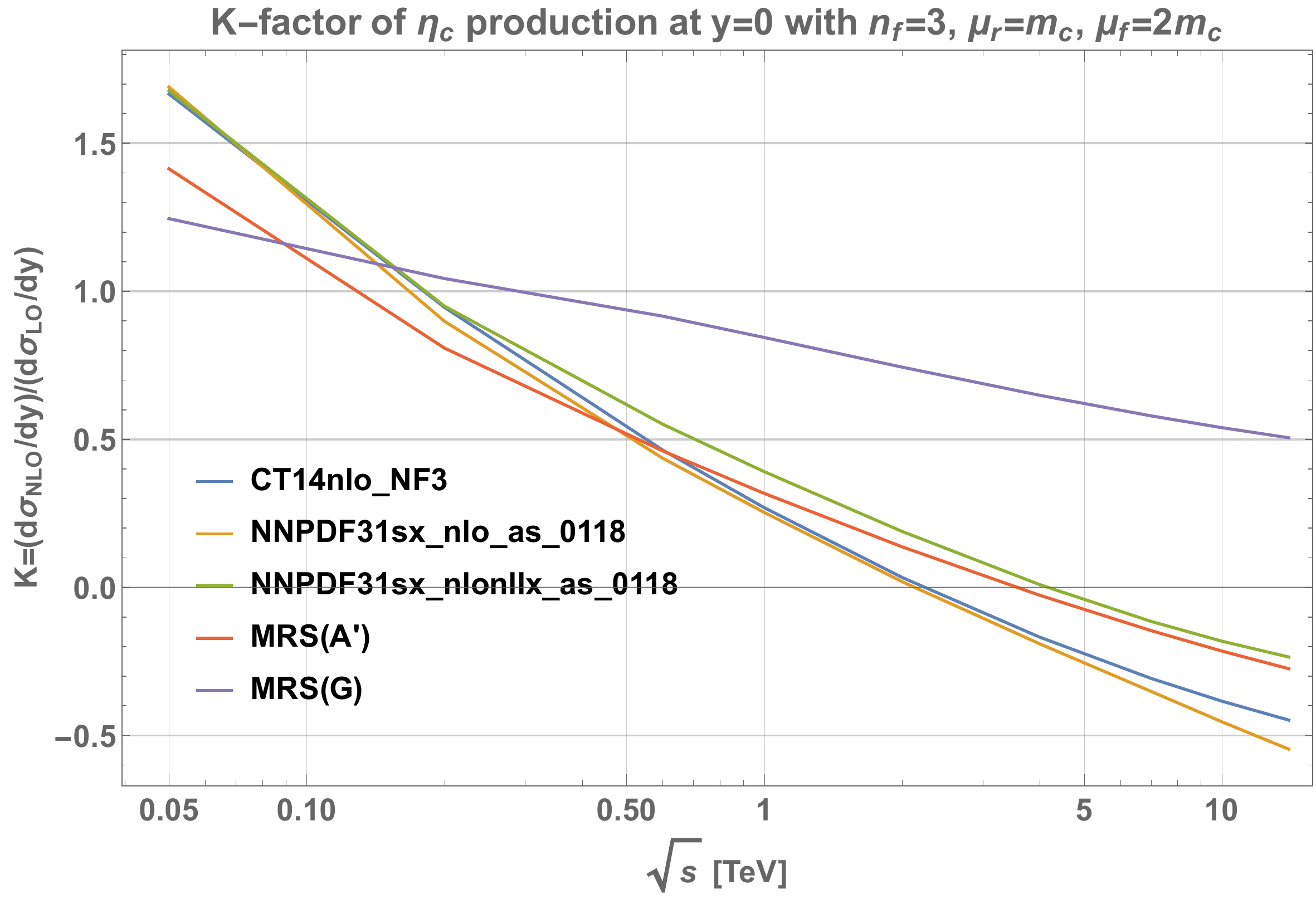}
	\setlength{\belowcaptionskip}{-6pt}
	\caption{$K$-factor at $y=0$ as a function of $\sqrt{s}$ and with different PDF choices. Default scale choice used $\mu_R=\mu_F=2m_c=3$GeV (left). Alternative scale choice used $\mu_R=m_c=1.5$GeV, $\mu_F=2m_c=3$GeV (right).}
	\label{fig:Ky0}
\end{figure}

Above, in Fig.~\ref{fig:Ky0}, we plot the evolution of the $K$-factor as a function of $\sqrt{s}$ for both scale configurations. One observes that the $K$-factor decreases as we approach higher energies. We can trace back this reduction of the $K$-factor to the real corrections that have negative contributions. In the case of the second scale configuration, for which QCD corrections are enhanced, the cross-section convoluted with each PDF set except one in our selection will give negative $K$-factors for $\sqrt{s}$ as low as $2$-$5$ TeV. As discussed above, these are very sensitive to $\lim_{z\to 0}{\hat{\sigma}}_{gg}$ that with the choice of $\mu_F=M=2m_c$ is negative. However the cross-section convoluted with MRS(G) stands out with its remarkable stable $K$-factor for both scale configurations. This is no surprise since MRS(G) has a very steep gluon PDF ($g(x)\sim 1/x^{1.30}$) and thus unlike the other PDF sets is much less sensitive to the partonic high-energy limit (see \EQ{eq:Schulerlimit}). As elaborated in the previous section, for MRS(G), it is therefore the 'positive' threshold contributions that dominate in these energy regions (albeit not in $\sqrt{s}\rightarrow \infty$ since $1.3<1.5$). We further note that at large $\sqrt{s}$ both NNPDFsxNLL and MRS(A') ($g(x)\sim 1/x^{1.14}$) have very similar $K$-factors.

As a side note, we mention here that NNPDFsxNLL has a steeper gluon PDF at low $x$ than NNPDFsx and therefore gives slightly better results than the latter one. This increase in low $x$ for the NLL extension can be explained by the fact that NLL contributions are slowing down DGLAP evolution. Therefore when the PDF is fitted to experimental data at larger factorisation scales, in order to accommodate a slower evolution, the initial parametrisation must become steeper such that the slower evolution matches the data.

Let us now briefly discuss the shape of the rapidity-differential cross-section at fixed energies $\sqrt{s}$. Due to lack of space, we give here only a qualitative description for the case of CT14. As before, we used the same two scale configurations. For $\mu_R=\mu_F=2m_c$, $d\sigma/dy$ at NLO starts to increase slightly with rapidity as we go to higher $\sqrt{s}$. As for the other scale choice, for which QCD corrections are enhanced, the curves become unphysical for large $\sqrt{s}$ where they start with a negative $d\sigma/dy$ at $y=0$ before rapidly getting positive at larger $y$.

Apart from constraining PDFs such that the cross-section is positive, we can thus certainly impose stronger constraints based on the two following criteria,
\begin{itemize}
\itemsep-0.3em
\item{$d\sigma/dy$ should increase with increasing energy $\sqrt{s}$ at any fixed rapidity $y$,}
\item{$d\sigma/dy$ should in general decrease with increasing rapidity $y$ at any fixed energy $\sqrt{s}$.}
\end{itemize}

\vspace*{-0.30cm}
\section{Conclusions}
\vspace*{-0.25cm}

Low-scale processes such as $\eta_c$ production depend crucially on the initial PDF parametrisation. If the gluon PDFs are not steep enough, real corrections can dominate and induce unphysical negative cross-sections. In addition to the positivity constraint of $d\sigma/dy\geq 0$, we have to impose that the cross-section increases with increasing energy at any fixed rapidity. Furthermore, since a NLO correction should not modify the rapidity-shape of the LO significantly, we could further make the constraint (less strong than the former one) that $d\sigma/dy$ decreases in general with increasing rapidity.

\vspace*{0.25cm}
{\setlength\parindent{0pt} {\bf Acknowledgements:} Some of the results presented are based on works made together with M.~G.~Echevarria, J.~P.~Lansberg, C.~Pisano and A.~Signori. The author thanks J.~P.~Lansberg and C.~Pisano for a critical reading of the manuscript and comments.}

\end{document}